\documentclass[twocolumn,showpacs,preprintnumbers]{revtex4}
\usepackage{amsmath}
\usepackage{graphicx}
\usepackage{dcolumn}
\usepackage{bm}
\usepackage{color}

\begin{document}
\title{The mixed quantum Rabi model}

\author{Liwei Duan$^{1}$, You-Fei Xie$^{1}$, and Qing-Hu Chen$^{1,2,*}$}

\address{
$^{1}$ Department of Physics and Zhejiang Province Key Laboratory of Quantum Technology and Device, Zhejiang University, Hangzhou 310027, China \\
$^{2}$ Collaborative Innovation Center of Advanced Microstructures,  Nanjing University,  Nanjing 210093, China}


\date{\today }

\begin{abstract}
The analytical exact solutions to the mixed quantum Rabi model (QRM) including
both one- and two-photon terms are found by using Bogoliubov operators.
Transcendental functions in terms of $4 \times 4$ determinants responsible
for the exact solutions are derived. These so-called $G$-functions with pole
structures can be reduced to the previous ones in the unmixed QRMs. The
zeros of $G$-functions reproduce completely the regular spectra. The
exceptional eigenvalues can also be  obtained by another transcendental
function. From the pole structure, we can derive two energy limits when the
two-photon coupling strength tends to the collapse point.
All energy levels only collapse to the lower one, which
diverges negatively. The  level crossings in the unmixed QRMs are
relaxed to avoided crossings in the present mixed QRM due to absence of
parity symmetry. In the weak two-photon coupling regime, the mixed QRM is
equivalent to an one-photon QRM with an effective positive bias, suppressed
photon frequency and enhanced one-photon coupling, which may pave a highly
efficient and economic way to access the deep-strong one-photon coupling
regime.
\end{abstract}


\maketitle

\section{Introduction}

The quantum Rabi model (QRM) describes the simplest and at the same time
most important interaction between a two-level system (or qubit) and a
single-mode bosonic cavity which is linear in the quadrature operators~\cite%
{braak2}. This model \ is a paradigmatic one in quantum optics for a long
time. \ It has been reactivated in the past decade, due to the progress in
many solid-state devices, such as the circuit quantum electrodynamics (QED)
\cite{Niemczyk,exp}, trapped ions \cite{ion1,ion2}, and quantum dots \cite{dot},
where the strong coupling even ultra-strong coupling has been realized. Here
we study a natural generalization of the QRM which exhibits both linear and
non-linear couplings between the qubit and the cavity, \emph{i.e.} the mixed QRM
having both one- and two-photon terms, with Hamiltonian
\begin{eqnarray}
H&=&-\frac{\Delta }{2}\sigma _{x}+\omega a^{\dagger }a \nonumber\\
&&+\sigma _{z}\left(
g_{1}\left( a^{\dagger }+a\right) +g_{2}\left[ \left( a^{\dagger }\right)
^{2}+a^{2}\right] \right) ,  \label{12p-rabimodel}
\end{eqnarray}%
where $\Delta $\ is the tunneling amplitude of the qubit, $\omega $ is the
frequency of cavity, $\sigma _{x,z}$ are Pauli matrices describing the
two-level system, $a$ ($a^{\dagger }$) is the annihilation (creation)
bosonic operator of the cavity mode, and $g_{1}$ ($g_{2}$) is the linear
(nonlinear) qubit-cavity coupling constant.

The nonlinear coupling appears naturally as an effective model for a
three-level system when the third (off-resonant) state can be eliminated.
The two-photon model has been proposed to apply to certain Rydberg atoms in
superconducting microwave cavities~\cite{Bertet,Brune}. Recently, a
realistic implementation of the two-photon QRM using trapped ions has been
proposed~\cite{Felicetti}. In the trapped ions, the atom-cavity coupling
could be tuned to the collapse regime.

The mixed QRM described by Eq.~(\ref{12p-rabimodel}) can also be implemented
in the proposal of the circuit QED~\cite{Felicetti1} if non-zero DC current
biases are applied. Using alternative methods, both linear and nonlinear
interaction terms can be present in different circuit QED setup by Bertet
\textit{et al.}~\cite{Bertet2, Bertet3}. Besides the one-photon process, the
two-photon process was also detected in the superconducting qubit and
oscillator coupling system \cite{tiefu}. It was shown
recently that a general Hamiltonian realized in the microwave driven ions
can be used to simulate the \ QRM with nonlinear coupling \cite{Jorge} by
chosing properly the time dependent phase and in a suitable interaction
picture. The combined linear and non-linear couplings can also be attained.
More recently, Pedernales \textit{et al.} proposed that a background of a ($%
1+1$)-dimensional black hole requires a QRM with both one- and two-photon terms
that can be implemented in a trapped ion for the quantum simulation of Dirac
particles in curved spacetime~\cite{Pedernales}. So the QRM with both one-
and two-photon couplings is not only a generic model in the circuit QED and
trapped ions, but also has applications in other realm of physics.

The unmixed QRMs, where either linear or nonlinear coupling is
present, have been extensively studied  for a few decades (for a review,
please refer to Refs.~\cite{treview,yuxi,ReviewF}). The solution based on
the well-defined $G$-function with pole structures was only found for
one-photon model by Braak \cite{Braak} in the Bargmann representation and
two-photon model by Chen \textit{et al.}~\cite{Chen2012} using Bogoliubov
transformations. These solutions have stimulated extensive research
interests in the exact solutions to the unmixed QRMs and their variants with
either one-photon~\cite{Zhong,Maciejewski,Chilingaryan,Peng,Wang, Fanheng}
or two-photon term~\cite{Trav,Trav1,Trav2,Maciejewski2,duan2016,Zhangyz,Zhangyz1,Lupo}. In the
literature, many analytical approximate but still very accurate results have
also been  given~\cite%
{Feranchuk,Irish,chenqh,zheng,chen2,chen3,yunbo,luo2,zhang,Zhiguo,cong,Casanova,PengJS}. In some
limits of model parameters, the dynamics and quantum criticality have
been studied exactly as well~\cite{plenio,hgluo,peng2}. In the unmixed QRMs, the
 parity symmetry is very crucial to get the analytical solution in the
closed system. Recently, the role of the parity symmetry has been
characterized in the excitation-relaxation dynamics of the system as a
function of light-matter coupling in the open systems~\cite{Malekakhlagh}

In the mixed QRM with both linear and nonlinear couplings, the parity
symmetry is however broken naturally, and the analytical solution thus
becomes more difficult \cite{xieyf}, compared to the unmixed models. In this
paper, we propose an analytical exact solutions to this mixed QRM. We derive a G-function by the Bogoliubov transformations, which can be reduced to
the previous G-functions for both one- and two-photon QRMs if either of the
couplings appears. We demonstrate that the derived $G$%
-function can really yield the regular spectra by checking with the
numerics. The exceptional eigenvalues are also given with the help of the
non-degeneracy property in this mixed model due to the absence of any
symmetry. Two kinds of formulae for the collapse points are  derived.
The avoided crossings are confirmed. The level collapse in the strong
two-photon coupling regime is also discussed. Finally, we study the
influence of mixed coupling by constructing  an equivalent one-photon QRM with
an effective positive bias where the photon frequency is suppressed and the
one-photon coupling is enhanced.

\section{Solutions within the Bogoliubov operators approach}

In the basis of spin-up and spin-down states, the Hamiltonian (\ref%
{12p-rabimodel}) can be transformed to the following matrix form in units of
$\omega =1$
\begin{widetext}
\begin{equation}
H=\left(
\begin{array}{ll}
a^{\dagger }a+g_{1}\left( a^{\dagger }+a\right) +g_{2}\left[ \left(
a^{\dagger }\right) ^{2}+a^{2}\right] & ~~~~~~~~-\frac{\Delta }{2} \\
~~~~~~~~-\frac{\Delta }{2} & a^{\dagger }a-g_{1}\left( a^{\dagger }+a\right)
-g_{2}\left[ \left( a^{\dagger }\right) ^{2}+a^{2}\right]%
\end{array}%
\right) .  \label{Hamiltonian}
\end{equation}
\end{widetext}

First, we perform Bogoliubov transformation
\begin{eqnarray}
A&=&S(r) D^{\dagger}(w) a D(w)S^{\dagger}(r) =ua+va^{\dagger }+w,\\
A^{\dagger }&=&S(r) D^{\dagger}(w) a^{\dagger} D(w)S^{\dagger}(r) =ua^{\dagger }+va+w,\nonumber
\end{eqnarray}%
to generate a new bosonic operator, where $S(r)$ is the squeezing operator and $D(w )$ is the displaced operator
\begin{equation*}
S(r)=e^{\frac{r}{2}(a^{2}-a^{\dag 2})}, ~~ D(w )=e^{w (a^{\dag }-a)},
\end{equation*}%
with $r=arc\cosh u$. If set
\begin{equation}
u=\sqrt{\frac{1+\beta }{2\beta }}, ~~ v=\sqrt{\frac{1-\beta }{2\beta }}, ~~ w=%
\frac{u^{2}+v^{2}}{u+v}g_{1},
\end{equation}%
with $\beta =\sqrt{1-4g_{2}^{2}}$, we have a simple quadratic form of one
diagonal Hamiltonian matrix element
\begin{equation}
H_{11}=\frac{A^{\dagger }A-v^{2}-w^{2}}{u^{2}+v^{2}}.  \label{H11_A}
\end{equation}%
The eigenstates of $H_{11}$ are the number states $\left\vert n\right\rangle _{A}$ which can be written in terms of the Fock states $\left\vert
n\right\rangle \ $ in original bosonic operator $a$ as
\begin{eqnarray}
\left\vert n\right\rangle _{A} =S(r)D^{\dag }(w)\left\vert n\right\rangle .
\end{eqnarray}%
Similarly, we can introduce another operator
\begin{eqnarray}
B&=&S^{\dagger}(r) D^{\dagger}(w^{\prime}) a D(w^{\prime})S(r) =ua-va^{\dagger }+w^{\prime},\\
B^{\dagger }&=&S^{\dagger}(r) D^{\dagger}(w^{\prime}) a^{\dagger} D(w^{\prime})S(r) =ua^{\dagger }-va+w^{\prime},\nonumber
\end{eqnarray}%
with
\begin{equation*}
w^{\prime }=\frac{u^{2}+v^{2}}{v-u}g_{1},
\end{equation*}%
which yields a simple quadratic form of the other diagonal Hamiltonian
matrix element%
\begin{equation}
H_{22}=\frac{B^{\dagger }B-v^{2}-w^{\prime 2}}{u^{2}+v^{2}}.  \label{H22_B}
\end{equation}%
Note that if $g_{2}=0$, we have $w=g_{1}$ and $w^{\prime}=-g_{1}$, which are exactly the same as those in the one-photon QRM~\cite{Chen2012}.
Similarly, the eigenstates of $H_{22}$ are  the number states $\left\vert
n\right\rangle _{B}$ which can be written in terms of the Fock states $\left\vert
n\right\rangle \ $ in original bosonic operator $a$ as
\begin{eqnarray}
\left\vert n\right\rangle _{B} = S^{\dag }(r)D^{\dag }(w^{\prime
})\left\vert n\right\rangle .
\end{eqnarray}

In terms of the Bogoliubov operator $A$, the Hamiltonian can be written as
\begin{equation}
H=\left(
\begin{array}{ll}
\frac{A^{\dagger }A-v^{2}-w^{2}}{u^{2}+v^{2}} & ~-\frac{\Delta }{2} \\
~~-\frac{\Delta }{2} & \;\;\;H_{22}^{\prime }%
\end{array}%
\right) ,
\end{equation}%
where
\begin{eqnarray}
H_{22}^{\prime } &=&\left( \left( u^{2}+v^{2}\right) +4g_{2}uv\right)
A^{\dag }A-2uv\left( \left( A^{\dag }\right) ^{2}+A^{2}\right)\nonumber\\
&& -2\left( u-v\right) ^{2}w\left( A^{\dag }+A\right) +h_{A},
\end{eqnarray}%
with%
\begin{equation*}
h_{A}=v^{2}+\left( u-v\right) ^{2}w^{2}\left( 1-2g_{2}\right) +2g_{1}\left(
u-v\right) w+2g_{2}uv.
\end{equation*}

In principle, the eigenfunctions of the Hamiltonian can be expanded in terms of the number states
of operator $A$%
\begin{equation}
\left\vert \psi \right\rangle _{A}=\sum_{n=0}^{\infty }\sqrt{n!}\left( \
\begin{array}{l}
e_{n}\left\vert n\right\rangle _{A} \\
f_{n}\left\vert n\right\rangle _{A}%
\end{array}%
\right) .  \label{wave_A}
\end{equation}%

Projecting the Schr$\overset{..}{o}$dinger equation onto $\left\vert
n\right\rangle _{A}$ gives
\begin{equation}
e_{n}=\frac{\Delta /2}{\frac{n-v^{2}-w^{2}}{u^{2}+v^{2}}-E}f_{n},
\label{coef1A}
\end{equation}%
\begin{widetext}
\begin{equation}
f_{n+2}=\frac{-\frac{\Delta }{2}e_{n}+\left[ \Omega \left( n,E\right) +h_{A}%
\right] f_{n}-2\left( u-v\right) ^{2}w\left( f_{n-1}+\left( n+1\right)
f_{n+1}\right) -2uvf_{n-2}}{2uv\left( n+1\right) \left( n+2\right) },
\label{coef2A}
\end{equation}
\end{widetext}
with
\begin{eqnarray*}
\Omega \left( n,E\right) &=&\left( u^{2}+v^{2}+4g_{2}uv\right) n-E \\
&=&\frac{\left( 1+4g_{2}^{2}\right) }{\beta }n-E.
\end{eqnarray*}%
This is actually a five terms recurrence relation for $f_{n}$. All
coefficients $e_{n}$ and $f_{n}$ for $n>1$ are determined in terms of $f_{0}$
and $f_{1}$ linearly.

In terms of the Bogoliubov operator $B$, the Hamiltonian can be written as
\begin{equation}
H=\left(
\begin{array}{ll}
\;\;\;H_{11}^{\prime } & ~-\frac{\Delta }{2} \\
~~-\frac{\Delta }{2} & \frac{B^{\dagger }B-v^{2}-w^{\prime 2}}{u^{2}+v^{2}}%
\end{array}%
\right) ,\nonumber
\end{equation}%
where
\begin{eqnarray}
H_{11}^{\prime } &=&\left( \left( u^{2}+v^{2}\right) +4g_{2}uv\right)
B^{\dag }B +2uv\left(
\left( B^{\dag }\right) ^{2}+B^{2}\right)\nonumber\\
&&-2\left( u+v\right) ^{2}w^{\prime }\left( B^{\dag }+B\right)+h_{B},\nonumber
\end{eqnarray}%
with
\begin{equation*}
h_{B}=v^{2}+\left( u+v\right) ^{2}w^{\prime 2} (1 + 2 g_2) -2g_{1}\left( u+v\right)
w^{\prime }+2g_{2}uv.\nonumber
\end{equation*}
We can express the eigenfunctions as
\begin{equation}
\left\vert {\psi}\right\rangle _{B}=\sum_{n=0}^{\infty }\sqrt{n!}\left( \
\begin{array}{l}
f_{n}^{\prime }\left\vert n\right\rangle _{B} \\
e_{n}^{\prime }\left\vert n\right\rangle _{B}%
\end{array}%
\right) .  \label{wave_B}
\end{equation}%
Similarly, we can get
\begin{equation}
e_{n}^{\prime }=\frac{\frac{\Delta }{2}}{\frac{n-v^{2}-w^{\prime 2}}{%
u^{2}+v^{2}}-E}f_{n}^{\prime },  \label{coef1B}
\end{equation}%
and the similar five terms recurrence relation for $f_{n}^{\prime }$.
Analogously, all coefficients $e_{n}^{\prime }$ and $f_{n}^{\prime }$ for $%
n>1$ are determined through $f_{0}^{\prime }$ and $f_{1}^{\prime }\ $
linearly.

Except for the crossing points in
the energy spectra, the eigenstates are nondegenerate. Two wavefunctions in terms of operators $A$ and $B$ correspond to the same eigenstate. Therefore, they should be  proportional with each other by a constant  $r$,
\begin{equation}
\sum_{n=0}^{\infty }\sqrt{n!}\left( \
\begin{array}{l}
e_{n}\left\vert n\right\rangle _{A} \\
f_{n}\left\vert n\right\rangle _{A}%
\end{array}%
\right) =r\sum_{n=0}^{\infty }\sqrt{n!}\left( \
\begin{array}{l}
f_{n}^{\prime }\left\vert n\right\rangle _{B} \\
e_{n}^{\prime }\left\vert n\right\rangle _{B}%
\end{array}%
\right) .  \label{twowave}
\end{equation}%
We will set $r=1$, because only ratios among $f_{0},f_{1},rf_{0}^{\prime }$
and $rf_{1}^{\prime }$ are relevant. In this case we can absorb $r$ into new
$f_{0}^{\prime }$ and $f_{1}^{\prime }$. Then we have
\begin{eqnarray}
\sum_{n=0}^{\infty }\sqrt{n!}e_{n}|n\rangle _{A} &=&\sum_{n=0}^{\infty }%
\sqrt{n!}f_{n}^{\prime }|n\rangle _{B},  \label{identity1} \\
\sum_{n=0}^{\infty }\sqrt{n!}f_{n}|n\rangle _{A} &=&\sum_{n=0}^{\infty }%
\sqrt{n!}e_{n}^{\prime }|n\rangle _{B}.  \label{identity2}
\end{eqnarray}

In the unmixed QRM, the well-defined $G$-functions can be derived by using
the lowest number state $\left\vert 0\right\rangle $ in the original Fock
basis for the one-photon model \cite{Braak,Chen2012}, and two lowest number
states $\left\vert 0\right\rangle $ and $\left\vert 1\right\rangle $ for the
two-photon model \cite{Chen2012,duan2016}. Here, we also project Eqs. (\ref%
{identity1}) and (\ref{identity2}) onto two original number states $%
\left\vert 0\right\rangle $ and $\left\vert 1\right\rangle $, and then
obtain the following 4 equations%
\begin{eqnarray}
G^{(0,0)} &=&\sum_{n=0}^{\infty }\sqrt{n!}\left[ f_{n}\langle 0|n\rangle
_{A}-e_{n}^{\prime }\langle 0|n\rangle _{B}\right] =0,  \label{G1} \\
G^{(0,1)} &=&\sum_{n=0}^{\infty }\sqrt{n!}\left[ f_{n}\langle 1|n\rangle
_{A}-e_{n}^{\prime }\langle 1|n\rangle _{B}\right] =0,  \label{G2} \\
G^{(1,0)} &=&\sum_{n=0}^{\infty }\sqrt{n!}\left[ e_{n}\langle 0|n\rangle
_{A}-f_{n}^{\prime }\langle 0|n\rangle _{B}\right] =0,  \label{G3} \\
G^{(1,1)} &=&\sum_{n=0}^{\infty }\sqrt{n!}\left[ e_{n}\langle 1 |n\rangle
_{A}-f_{n}^{\prime }\langle 1|n\rangle _{B}\right] =0.  \label{G4}
\end{eqnarray}%
They form $4$ sets of linear homogeneous equations with $4$ unknown
variables $f_{0},f_{1},f_{0}^{\prime },$and $f_{1}^{\prime }$. Nonzero
solutions require the vanishing of the following $4\times 4$ determinant
\begin{equation}
G(E)=\left\vert G_{i,j}\right\vert =0,  \label{G-func}
\end{equation}%
where elements $G_{i,j}s$ are just coefficients before $f_{0},f_{1},f_{0}^{%
\prime },$and $f_{1}^{\prime }$ in Eqs. (\ref{G1})-(\ref{G4}).

Eq.~(\ref{G-func}) is just the $G$-function of the present mixed QRM. Its zeros thus give all regular eigenvalues of
the mixed QRM, which in turn give the eigenstates using Eq.~(\ref{wave_A})
or Eq.~(\ref{wave_B}). Note from the coefficients in Eqs.~(\ref{coef1A}) and
(\ref{coef2A}) that this $G$-function is a well-defined transcendental
function. Thus analytical exact solutions have been formally found. In the
next section, we will employ it to analyze the characteristics of the
spectra.

Note that for the one-photon QRM ($g_2=0$), the parity symmetry leads to%
\begin{eqnarray*}
e_{n}^{\prime } =\pm \left( -1\right) ^{n}e_{n}, ~~
f_{n}^{\prime } =\pm \left( -1\right) ^{n}f_{n}.
\end{eqnarray*}%
then Eq.~(\ref{G1}) becomes
\begin{equation}
G^{(0,0)}\left( E\right) = \sum_{n=0}^{\infty }\sqrt{n!}\left[ f_{n}\langle
0|n\rangle _{A}\mp \left( -1\right) ^{n}e_{n}\langle 0|n\rangle _{B}\right],
\label{G_1p}
\end{equation}%
which is just the $G$-function of one-photon QRM ($g_2=0$)~\cite{Braak}.
Similarly, Eq.~(\ref{G1}) (Eq.~(\ref{G2})) can be reduced to the previous
ones~\cite{Chen2012,duan2016} of the two-photon QRM ($g_1=0$) in the subspace
with even (odd) bosonic number.

In various unmixed QRMs, such as one-photon~\cite{Braak,Chen2012}, two-photon~\cite{duan2016} or two-mode~\cite{liwei2015} QRM, the coefficients of the eigenstates satisfy a three-term recurrence relation which can be achieved by performing the Bogoliubov transformation.
All these  $G$-functions with explicit pole structures  have been  summarized in Eq. (27) of Ref.~\cite{liwei2015}. But in the mixed model,
the Hilbert space cannot be separated into invariant subspaces due to the lack of parity symmetry, so the
recurrence relation of the coefficients $\{f_n\}$ is of higher order, as seen in Eq. (\ref{coef2A}). A
similar behavior also happens in the Dicke model \cite{Braak2013, Heshu}.
The possible reason is that the symmetry does not suffice to label each state
uniquely,  indicating that mixed QRM is non-integrable
according to Braak's criterion for quantum intergrability \cite{Braak}.

\section{Exact spectra}

\subsection{Regular spectra}

\begin{figure}[tbp]
\centering
\includegraphics[width=8.5cm]{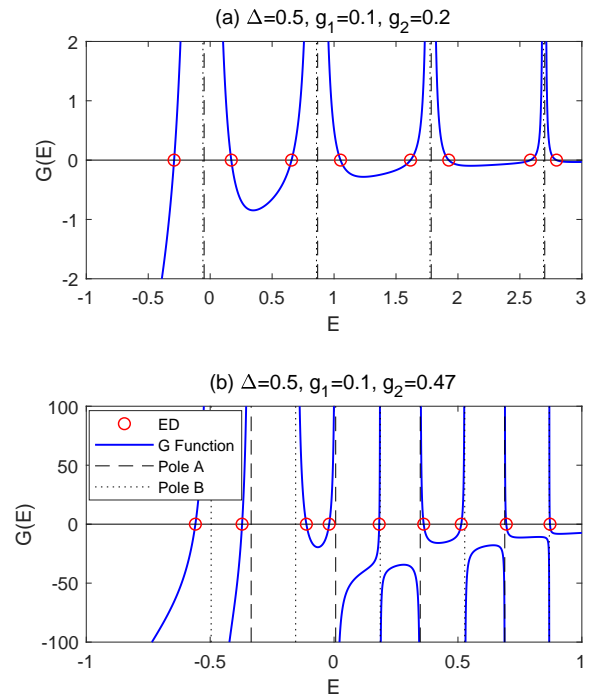}
\caption{ $G$-curves for $\Delta =0.5,g_{1}=0.1$. (a) $g_{2}=0.2$, (b) $g_{2}=0.47$. The blue solid lines denote the $G$-curves. The dash (dot) lines denote poles associated with $A$ ($B$) operator. The red circles denote the results calculated by numerical diagonalization.}
\label{G-function}
\end{figure}
To show the validity of the $G$-function (\ref{G-func}), we first check with
independent numerics. The $G$-curves as a function of $E$ for $\Delta =0.5$, $g_{1}=0.1$, $g_{2}=0.2$ and $0.47$
are depicted in Fig. \ref{G-function}. We find that the zeros of $G$%
-function indeed yield the true eigenvalues by comparing with the numerical
diagonalization in truncated Hilbert spaces with sufficiently high dimension.

\begin{figure}[tbp]
\centering
\includegraphics[width=8.5cm]{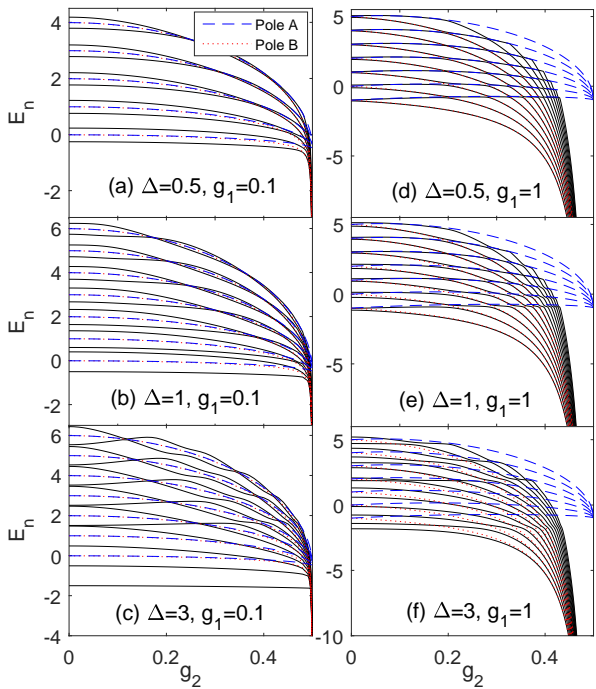}
\caption{ Energy spectra as a function of $g_2$. The solid black lines denote the energy spectra obtained from the $G$-function. The blue dash (red dot) lines denote poles associated with $A$ ($B$) operator.}
\label{spectra}
\end{figure}

Then we plot the energy spectra calculated by the $G$-function (\ref{G-func}) in Fig. \ref%
{spectra}. Checking with numerics, our $G$-function reproduces completely
all eigenvalues of the present mixed model.
When $g_{2}$ is close to $1/2$, the energy spectra collapse to negative infinity.
 The parity symmetry in this mixed QRM is lacking, so in principle, the energy
degeneracy should be relieved, and level crossings should be absent. However, as shown in Fig. \ref{spectra} (a)-(c), it seems
that some crossings still occur  for
the small $g_{1}$. It will be shown later that these "crossing" can
be actually discerned as avoided crossings.

\subsection{Pole structure and collapse}

From Eqs. (\ref{coef1A}) and (\ref{coef1B}), we can find two kinds of poles
associated with the $A$ and $B$ operators respectively, which will lead to the divergency
of the recurrence relations,
\begin{eqnarray}
E_{n}^{\mathrm{(pole\_A)}} &=&\beta n-\frac{1-\beta }{2}-\frac{g_{1}^{2}}{1+2g_{2}},
\label{poleA} \\
E_{n}^{\mathrm{(pole\_B)}} &=&\beta n-\frac{1-\beta }{2}-\frac{g_{1}^{2}}{1-2g_{2}}.
\label{poleB}
\end{eqnarray}%
With the same $n$, the difference of two poles is independent of $n$.

\begin{equation*}
\Delta E^{\mathrm{(p)}}=\frac{g_{1}^{2}}{1-2g_{2}}-\frac{g_{1}^{2}}{1+2g_{2}}%
=g_{1}^{2}\frac{4g_{2}}{\beta ^{2}}.
\end{equation*}%
In the limit of $g_{2}\rightarrow 1/2,$ $\beta \rightarrow 0$, all $%
E_{n}^{\mathrm{(pole\_A)}}$ are squeezed into a single finite value $\ -\frac{1}{2}\left( 1+g_{1}^{2}\right) $, while
all $E_{n}^{\mathrm{(pole\_B)}}$ diverge to $-\infty $. It seems that there are two kinds
of collapse energies. But actually, all energy levels tend to $B$-poles only, namely $-\infty $, if
$g_{2} \rightarrow 1/2$,  as shown in Fig.~\ref{spectra}. This spectral characteristics  is quite different from the two-photon QRM where the energy spectra
collapse to the finite value~\cite{Ng,Felicetti,duan2016}. For the mixed QRM, the divergence of the eigenenergies to negative infinity for $g_2 \rightarrow 1/2$ suggests some underlying unphysicality, which deserves further studies.
The energies of the high excited states cross
the pole A curves and then asymptotically converge to the pole B,
which lead to exceptional solutions.

\subsection{Exceptional solutions}

\begin{figure}[tbp]
\centering
\includegraphics[width=9cm]{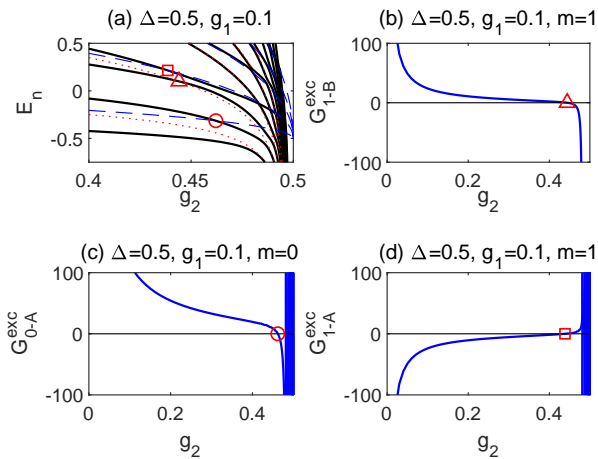}
\caption{(a) Enlarged view of energy spectra in Fig. \ref{spectra} (a). The solid black lines denote the energy spectra obtained from the $G$-function. The blue dash (red dot) lines denote poles associated with $A$ ($B$) operator. The triangle denotes the first intersecting point of the energy levels and the $1$-st pole line associated with the $B$ operator. The circle (square) denotes the intersecting point of the energy level and the $0$-th ($1$-st) pole line associated with the $A$ operator. (b)
The B-type exceptional eigenvalue for $m=1$. The A-type
exceptional eigenvalues for (c) $m=0$ and (d) $m=1$. $\Delta =0.5$, $g_{1}= 0.1$.}
\label{exceptional}
\end{figure}

As shown in  Fig. \ref{exceptional} (a), most energy level curves pass through the pole curves on the
way to $g_{2}=1/2$, which results in so-called exceptional solutions. They can
be located in the following way.

At the intersecting point of the energy levels and the $m$-th pole line
associated with the $A$-operator (\ref{poleA}), the coefficient $f_{m}$ must
vanish so that the pole is lift. Otherwise, the coefficient $e_{m}$ would
diverge due to zero denominator in Eq. (\ref{coef1A}). In the unmixed QRM, $%
f_{m}=0$ can uniquely yield the necessary and  sufficient condition for the
occurrence of the exceptional solution. But here it is not that case,
because $f_{m}$ depends on two initial variable $f_{0},f_{1}$, and cannot be
determined uniquely. The corresponding coefficient $e_{m}$ should be finite
and can be regarded as an unknown variable. In
all summations in Eqs.~(\ref{G1})$-$(\ref{G4}), the $m$-th terms should be
treated specially, i.e. let $f_{m}$ be $0$ and $e_{m}$ be a new variable. By
the recurrence relation (\ref{coef2A}), we can add a new equation for this
case
\begin{widetext}
\begin{equation}
f_{m}=\frac{-\frac{\Delta }{2}e_{m-2}+\left[ \Omega \left( m-2,E\right) +h_{A}%
\right] f_{m-2}-2\left( u-v\right) ^{2}w\left( f_{m-3}+\left( m-1\right)
f_{m-1}\right) -2uvf_{m-4}}{2uv m\left( m-1\right)  }=0.  \label{new_eq}
\end{equation}%
\end{widetext}
So for the exceptional solution, we have a set of linear homogeneous
equations (Eqs. (\ref{G1})-(\ref{G4}) and (\ref{new_eq})) with $5$ unknown variables $f_{0}$, $f_{1}$, $f_{0}^{\prime}$, $f_{1}^{\prime }$ and $e_{m}$ for $m \ge 2$. While for $0 \le m<2$, $f_m=0$%
, we have only $4$ unknown variables $f_{1-m}$, $f_{0}^{\prime }$, $f_{1}^{\prime }$ and $%
e_{m}$,  which can be determined by solving another set of linear homogeneous
equations (Eqs. (\ref{G1})-(\ref{G4})). Nonzero solution requires the vanishing of the  $%
5\times 5$ ($4\times 4$) determinant  whose elements are just coefficients before $f_{0}$, $f_{1}$, $f_{0}^{\prime}$, $f_{1}^{\prime }$ and $e_{m}$ in Eqs. (\ref{G1})-(\ref{G4}) and (\ref{new_eq}) ($f_{1-m}$, $f_{0}^{\prime }$, $f_{1}^{\prime }$ and $%
e_{m}$ in Eqs. (\ref{G1})-(\ref{G4})) for $m \ge 2$ ($0 \le m<2$),
\begin{equation}
G_{m-A}^{\mathrm{exc}}\left( \Delta ,g_{1},g_{2}\right) =0.  \label{except_A}
\end{equation}%
We call this function as exceptional $G$-function. Here the energy is not an
explicit variable, but determined by Eq. (\ref{poleA}). The $m$-th exceptional solution associated with the $B$ operator can be
detected in the same way by zero of an exceptional $G$-function
\begin{equation}
G_{m-B}^{\mathrm{exc}}\left( \Delta ,g_{1},g_{2}\right) =0.  \label{except_B}
\end{equation}%

With the help of the exceptional $G$-function, we can determine the intersecting points of the energy levels and the pole curves as shown in Fig. \ref{exceptional} (a) for $%
\Delta =0.5,g_{1}=0.1$. The $G^{\mathrm{exc}}$ curves associated with $A$%
-pole as a function of $g_{2}$ for $m=0$ and $1$ are shown in Fig. \ref{exceptional} (c) and (d) respectively, and  the $G^{\mathrm{exc}}$ curve associated with
$B$-pole for $m=1$ is shown in Fig. \ref{exceptional} (b). The detected
exceptional solutions in the $G^{\mathrm{exc}}$ curves  are marked by
the same symbols as those in the enlarged spectra graph. One can find many zeros for $G^{%
\mathrm{exc}}$ curve associated with $A$-pole in Fig. \ref{exceptional} (c) and (d), which are corresponding to
intersecting points of  energy levels and the $m$-th $A$-pole curves as displayed in Fig. \ref{exceptional} (a). For $G^{\mathrm{exc}}$ curve associated with $B$-pole, there is only one zero
for $m=1$ as exhibited in Fig. \ref{exceptional} (b), also consistent with the single intersecting point shown in Fig. \ref{exceptional} (a). No exceptional
solution exists even for $B$-pole with $m=0$. All $g_{2}$ obtained from $G^{\mathrm{exc}}=0$ can find
their corresponding intersecting points in the energy spectra exactly.

Now we can judge whether it is the true level crossing or avoided crossing
in the Fig. \ref{spectra} (a)-(c).
Around this regime, we have not found any exceptional solutions, indicating
that the energy level cannot intersect with the pole curves. So although two
energy levels are very close but blocked off by two pole curves with
difference $\Delta E^{\mathrm{(p)}}\propto $ $g_{1}^{2}$, they neither collide nor
cross with each other. It is actually avoided crossing. For small $g_{1}$, $%
\Delta E^{\mathrm{(p)}}$ is very small, so it looks like a "level crossing" as depicted
in   Fig. \ref{spectra} (a)-(c). For large $g_{1}$, the avoided
crossing is quite clear, as shown in   Fig. \ref{spectra} (d)-(f).
Actually, these avoided crossings are just remnants of the traces of the
doubly degeneracy in the unmixed model, which is relieved in the mixed model.

\section{Effect of the mixed couplings}

\begin{figure}[tbp]
\centering
\includegraphics[width=8.5cm]{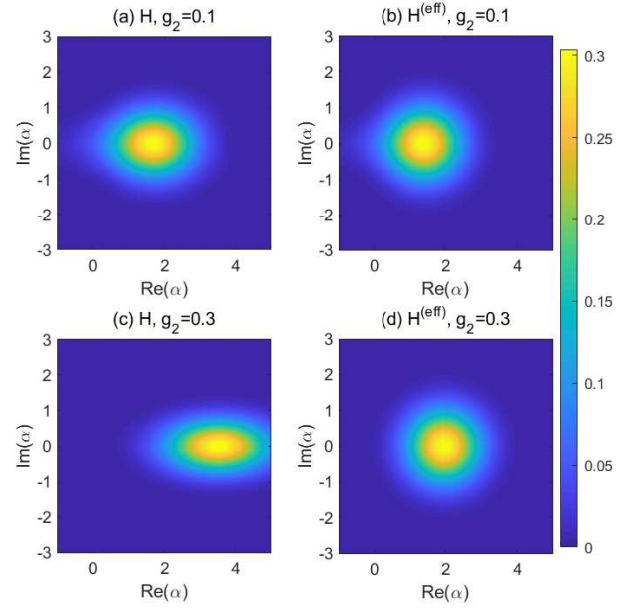}
\caption{The Wigner function of the ground state  at $\Delta=1$, $g_1=1$, with (a)-(b) $g_2=0.1$ and (c)-(d) $g_2=0.3$. Left column is calculated from the original Hamiltonian $H$, and the right column is calculated from the effective Hamiltonian $H^{(\mathrm{eff})}$.}
\label{wigner}
\end{figure}

\begin{figure}[tbp]
\centering
\includegraphics[width=7.5cm]{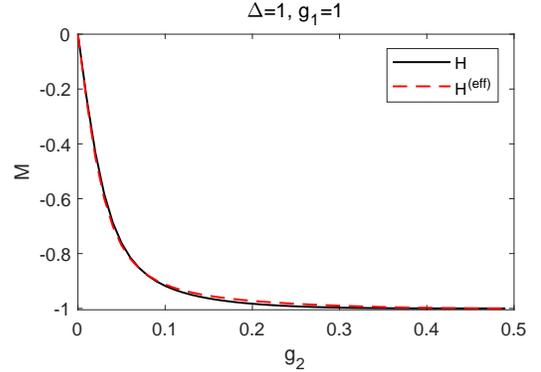}
\caption{The ground-state magnetization $M$ as a function of $g_2$ for $\Delta=1$, $g_1=1$. Results for the full mixed model (\ref{12p-rabimodel}) are denoted by black solid lines, agreeing well with those by effective Hamiltonian (\ref{Heff}) denoted by red dashed lines.}
\label{sz_g2}
\end{figure}

In the mixed QRM, if we combine Eqs.~(\ref{H11_A}) and (\ref{H22_B}),
the Hamiltonian (Eq.~(\ref{Hamiltonian})) can be written as
\begin{equation}
H=\left(
\begin{array}{ll}
\frac{A^{\dagger }A-v^{2}-w^{2}}{u^{2}+v^{2}} & ~~~~~-\frac{\Delta }{2} \\
~~-\frac{\Delta }{2} & \;\;\;\frac{B^{\dagger }B-v^{2}-w^{\prime 2}}{%
u^{2}+v^{2}}%
\end{array}%
\right) ,
\end{equation}%
which can be reorganized and  separated into three terms:
\begin{eqnarray}
H &=&H_{0}+\frac{\epsilon ^{(\mathrm{eff})}}{2}\sigma _{z} -  \frac{1 - \beta}{2} \mathbf{I},
\end{eqnarray}
where $\mathbf{I}$ is a unit matrix, and
\begin{eqnarray}
H_{0} &=&\left(
\begin{array}{ll}
\beta A^{\dagger }A-\frac{g_{1}^{2}}{\beta ^{2}} &
~~~~~~-\frac{\Delta }{2} \\
~~~~~-\frac{\Delta }{2} & \;\;\;\beta B^{\dagger }B-%
\frac{g_{1}^{2}}{\beta ^{2}}%
\end{array}%
\right) ,  \label{H_0} \\
\epsilon ^{(\mathrm{eff})} &=&\frac{4g_{2}}{1-4g_{2}^{2}}g_{1}^{2}.
\end{eqnarray}%
An effective bias $\epsilon^{\mathrm{(eff)}}$ appears naturally, as well as a
total energy shift $(1 - \beta)/2$.
Recalling the one-photon QRM~\cite{Chen2012},
\begin{eqnarray}
H_{\mathrm{1P}} = -\frac{\Delta}{2} \sigma_x + \omega a^{\dagger} a + g_1 \sigma_z (a^{\dagger} + a),\label{H_1p}
\end{eqnarray}
this Hamiltonian can be expressed with a new set of bosonic operators $P = D^{\dagger}(g_1) a D(g_1) = a + g_1$
and $Q = D(g_1) a D^{\dagger}(g_1) = a - g_1$ as
\begin{eqnarray}
H_{\mathrm{1P}} = \left(
\begin{array}{ll}
\omega P^{\dagger }P-\frac{g_{1}^{2}}{\omega} &
~~~~~~-\frac{\Delta }{2} \\
~~~~~-\frac{\Delta }{2} & \;\;\;\omega Q^{\dagger }Q-%
\frac{g_{1}^{2}}{\omega}%
\end{array}%
\right) .\label{H_1p_AB}
\end{eqnarray}
Comparing Eqs.~(\ref{H_0}) and (\ref{H_1p_AB}), we can introduce an effective photon frequency
$\omega^{\mathrm{(eff)}}$ and an effective one-photon coupling strength $g_1^{\mathrm{(eff)}}$,
\begin{eqnarray}
\omega^{\mathrm{(eff)}} &=& \beta,\nonumber\\
g_1^{\mathrm{(eff)}} &=& \frac{g_1}{\sqrt{\beta}},\nonumber
\end{eqnarray}
and rewrite $H_0$ as
\begin{eqnarray*}
H_{0} = -\frac{\Delta}{2} \sigma_x + \omega^{\mathrm{(eff)}} a^{\dagger} a + g_1^{\mathrm{(eff)}} \sigma_z (a^{\dagger} + a).
\end{eqnarray*}
Therefore, we construct an effective one-photon QRM to describe the mixed one, which provides
a more intuitional description of the influence of the mixed coupling,
\begin{eqnarray}
H^{\mathrm{(eff)}} &=& \frac{\epsilon ^{(\mathrm{eff})}}{2}\sigma _{z}
-\frac{\Delta}{2} \sigma_x  -  \frac{1 - \beta}{2}\nonumber\\
 &&+ \omega^{\mathrm{(eff)}} a^{\dagger} a + g_1^{\mathrm{(eff)}} \sigma_z (a^{\dagger} + a).\label{Heff}
\end{eqnarray}
Comparing $P$ ($Q$) with $A$ ($B$), the mainly difference is the lack of squeezing operator $S(r)$. The two-photon interaction  leads to the squeezed field state~\cite{PengJS,Felicetti1}, which can be well captured by the squeezing operator~\cite{Chen2012,duan2016,Lupo}. Therefore, the squeezing operator is introduced explicitly to deal with the mixed QRM, but it is not necessary for the one-photon model.
The definitions of $P$ ($Q$) in Eq.~(\ref{H_0})
is equivalent with that of $A$ ($B$) in Eq.~(\ref{H_1p_AB})  only if $g_{2}=0$, so it is hard to use the
effective Hamiltonian to deal with the strong two-photon coupling regime, especially the bosonic part due to the intense squeezing effect. We calculate the Wigner function $W(\alpha, \alpha^*)$ of the ground state~\cite{qutip1,qutip2} in Fig.~\ref{wigner}, which describes the probability distribution of the bosonic field in the phase space. When $g_2=0.1$, the differences of the Wigner functions calculated from $H$ and $H^{(\mathrm{eff})}$ are negligible. However, It is shown in Fig.~\ref{wigner} (c) that the squeezing effect becomes apparent for $H$ when $g_2=0.3$. The effective Hamiltonian can hardly describe the squeezing effect as demonstrated  in Fig.~\ref{wigner} (d).
Nevertheless, it shed light on the analysis of strong coupling case, especially the properties of the two-level system. The ground-state magnetization $M=\langle \psi_{GS} |\sigma_z |\psi_{GS}\rangle$ calculated from $H^{(\mathrm{eff})}$ is in good agreement with that calculated from $H$ even in strong two-photon coupling regime, as shown in Fig.~\ref{sz_g2}. When $g_{2}$
tends to  $1/2$, the effective bias $\epsilon ^{(\mathrm{eff})}$ will tend to
infinity, and the two-level system will prefer to stay in the lower level as indicated by the ground-state magnetization $M \rightarrow -1$  in Fig.~\ref{sz_g2}.  Therefore, the energy contributed by $\epsilon^{(\mathrm{eff})} \sigma_z / 2$ would be negative infinite, which is one of the reason for the negative divergence of the eigenenergies in the mixed QRM in the limit of $g_2 \rightarrow 1/2$, as observed
in Fig.~\ref{spectra}. What is more, with the
increase of $g_{2}$, the effective photon frequency $\omega ^{(\mathrm{eff})}
$ will decrease while the effective coupling $g_{1}^{(\mathrm{eff})}$ will
increase, which might provide a novel and economic way to reach deep-strong
one-photon coupling regime.

\begin{figure}[tbp]
\centering
\includegraphics[width=8.5cm]{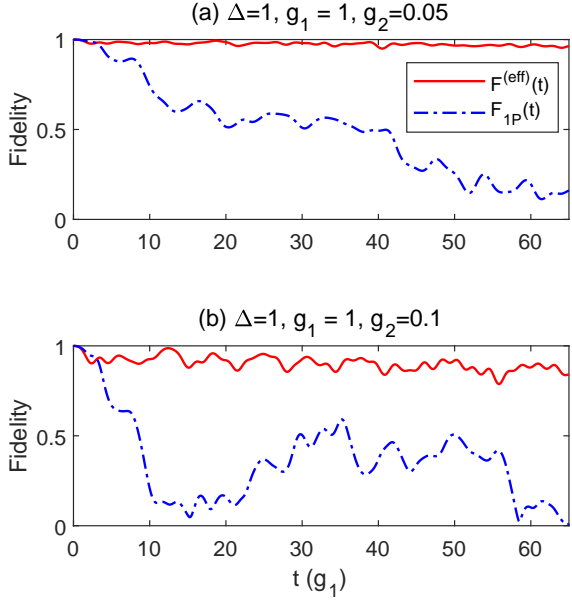}
\caption{The fidelity of $H^{(\mathrm{eff})}$ (red sold line) and $H_{\mathrm{1P}}$ (blue dash-dot line) as a function of $t$ for $%
\Delta=1$, $g_1=1$, (a) $g_2=0.05$ and (b) $g_2=0.1$ .}
\label{dynamics}
\end{figure}
To further demonstrate the accuracy of the effective Hamiltonian, we calculate the dynamics of fidelity as shown in Fig.~\ref{dynamics}. The fidelity is defined as the overlap of the wavefunctions $|\psi^{(\mathrm{eff})}(t)\rangle$ obtained from  $H^{(\mathrm{eff})}$ (Eq. (\ref{Heff})) and $|\psi(t)\rangle$ obtained from  $H$ (Eq. (\ref{12p-rabimodel})), namely $F^{(\mathrm{eff})}(t)=|\langle \psi^{(\mathrm{eff})}(t) |\psi(t)\rangle|$, which can be used to judge  how accurately the state of the effective Hamiltonian reproduces that of the original Hamiltonian. The initial states are $|\uparrow\rangle |0\rangle$ for both of them. The
fidelity of the unbiased one-photon QRM, namely $H_{\mathrm{1P}}$ (Eq. (\ref{H_1p})), is also
presented for comparison, \textit{i. e.} $F_{1P}(t)=|\langle \psi_{1P}(t) |\psi(t)\rangle|$. When $g_2$ is as small as $0.05$, the corresponding effective bias reaches $%
\epsilon^{(\mathrm{eff})}\simeq 0.202$. This effective bias
is large enough to play a significant
role in  the evolution of fidelity, as clearly seen  in the  Fig.~\ref{dynamics} (a). The fidelity of $H^{(\mathrm{eff})}$ tends to  one, which is a strong evidence of the equivalence between $|\psi^{(\mathrm{eff})}(t)\rangle$  and $|\psi(t)\rangle$. The fidelity of $H_{\mathrm{1P}}$ is much smaller,  indicating that it deviates from the original Hamiltonian significantly.
When we further increase $g_2$, the effective Hamiltonian still gives considerably good results while the deviation of $H_{\mathrm{1P}}$ becomes more obvious. The fidelity of $H^{(\mathrm{eff})}$ drops slightly in the long time, which is mainly due to the error accumulation. The results of fidelity confirm the limitation of the effective Hamiltonian in dealing with the strong two-photon coupling and long-time limits.

\begin{figure}[tbp]
\centering
\includegraphics[width=8.5cm]{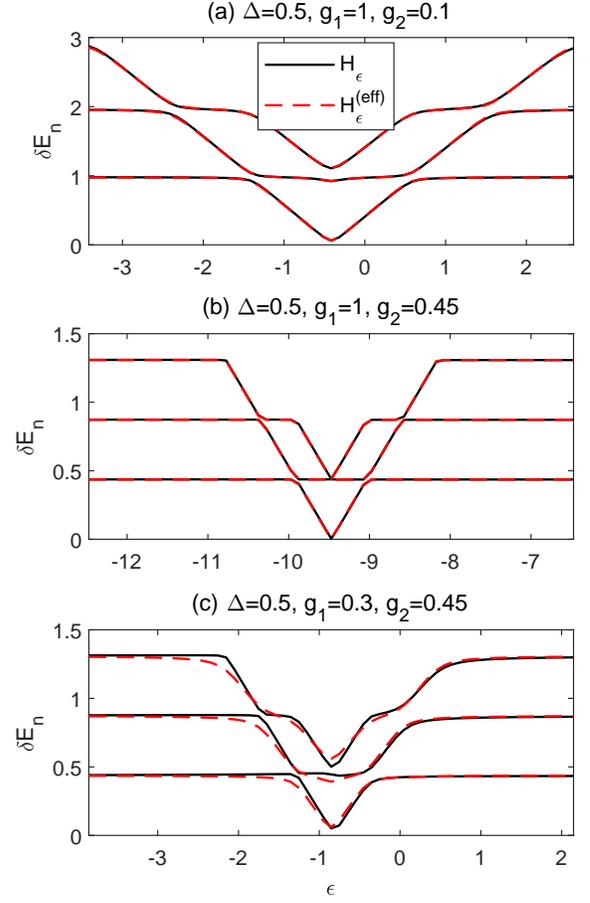}
\caption{The energy difference $\protect\delta E_{n}$ as a
function of $\protect\epsilon $ for (a) $\Delta =0.5$, $g_{1}=1$, $g_{2}=0.1$, (b) $\Delta =0.5$, $g_{1}=1$, $g_{2}=0.45$, and (c) $%
\Delta =0.5$, $g_{1}=0.3$, $g_{2}=0.45$. Results calculated
from $H_{\protect\epsilon }$, and $H_{\protect\epsilon }^{(\mathrm{eff})}$
are marked with black solid and red dash lines respectively.}
\label{dE_tot}
\end{figure}
The mixed QRM can be realized in the experiment by coupling the
flux qubit to the plasma mode of its DC-SQUID detector~\cite{Bertet2}.
We expect that the effective Hamiltonian can be employed to explain the
experimental results. One of the most widely measured quantity in experiments
 is the transmission spectrum~\cite{exp, Yoshihara,Forn2}.
The transmission spectrum, \textit{i.e.} $\delta E_{n}=E_{n}-E_{0}$,
from both original full model (\ref{12p-rabimodel}) and the effective model (\ref{Heff}) are shown in Fig.~\ref{dE_tot}.
We introduce an additional bias $\epsilon
\sigma _{z}/2$ which is originated from the an externally applied
magnetic flux in circuit QED system, and Eqs.~(\ref{12p-rabimodel}) and (\ref{Heff}) become
\begin{eqnarray}
H_{\epsilon } &=&\frac{\epsilon }{2}\sigma _{z}-\frac{\Delta }{2}\sigma
_{x}+a^{\dagger }a  \notag \\
&&+\sigma _{z}\left( g_{1}\left( a^{\dagger }+a\right) +g_{2}\left[ \left(
a^{\dagger }\right) ^{2}+a^{2}\right] \right) ,  \label{H_e} \\
H_{\epsilon }^{(\mathrm{eff})} &=&\frac{\epsilon +\epsilon ^{(\mathrm{eff})}%
}{2}\sigma _{z}-\frac{\Delta }{2}\sigma _{x}  \nonumber\\
&&+\omega ^{\mathrm{(eff)}}a^{\dagger }a+g_{1}^{\mathrm{(eff)}}\sigma
_{z}(a^{\dagger }+a).\label{H_e_eff}
\end{eqnarray}%
Therefore, the total bias in the effective Hamiltonian is ($\epsilon ^{(%
\mathrm{eff})}+\epsilon $). It is obvious from Fig.~\ref{dE_tot} (a) that the effective Hamiltonian can  capture
the effects of two-photon coupling in the weak coupling regime very well. Even for strong
two-photon coupling $g_{2}=0.45$, the effective Hamiltonian (\ref{H_e_eff})
still provides quite accurate energy structure and almost captures all
features of the mixed QRM, as shown in  Fig.~%
\ref{dE_tot} (b)-(c). With the decrease of $g_1$, the deviation appears gradually
because the two-photon interaction becomes dominated.  One should also note that the
energy differences of the effective Hamiltonian (Eq.~(\ref{H_e_eff})) is symmetry about $\epsilon
=-\epsilon ^{(\mathrm{eff})}$, which is different from that of the original one
(Eq.~(\ref{H_e})), as shown in Fig.~\ref{dE_tot} (c). For the effective
Hamiltonian with an additional bias, we can  easy confirm that only the
absolute value of $(\epsilon +\epsilon ^{(\mathrm{eff})})$ affects the
eigenenergies, as well as the energy differences. For the mixed QRM, this
symmetry is broken due to the two-photon interaction term. Therefore,
the asymmetry in the transmission spectrum can be regarded as a signature of
the mixed one- and two-photon couplings. Far  from the symmetry point $%
\epsilon =-\epsilon ^{(\mathrm{eff})}$, the energy difference tends to be a
multiple of the photon frequency. The energy difference decreases with the
increase of $g_{2}$, which can be explained by the suppressed effective
photon frequency $\omega^{(\mathrm{eff})}$.

\begin{figure}[tbp]
\centering
\includegraphics[width=8cm]{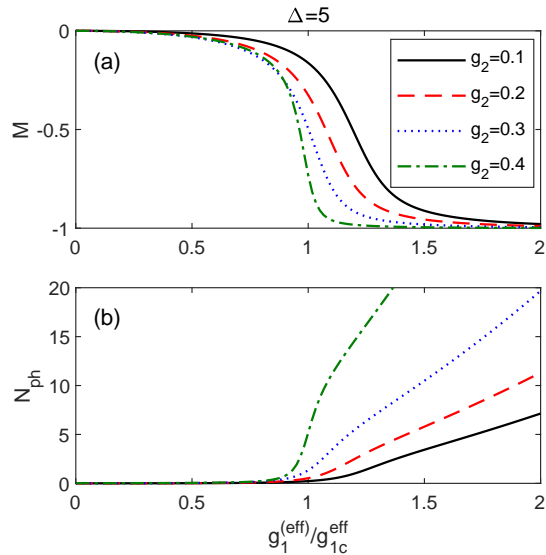}
\caption{ (a) The magnetization $M$  and (b) photon
number $N_{\mathrm{ph}}$  calculated from $H$ as a function of $%
g_1^{(\mathrm{eff})}$ for $\Delta =5$, $g_2=0.1$ (black solid), $0.2$ (red dash), $0.3$ (blue dot) and $0.4$ (green dash-dot lines). }
\label{Sz_Nph}
\end{figure}
Recently, the quantum phase transition of one-photon QRM in
$\Delta/\omega\rightarrow\infty$ has drawn much attention~\cite{plenio,hgluo}.
 The magnetization $M = \langle \sigma_z \rangle$ serves as an order parameter
 which changes from zero in the normal phase to nonzero
 in the superradiant phase when $g_1$ crosses the critical point
 $g_{1c} = \sqrt{\Delta \omega} / 2$. Above $g_{1c}$,
 photons are extremely activated as well.
The ground-state magnetization $M$ and the photon
number $N_{\mathrm{ph}} = \langle a^{\dagger} a \rangle$ calculated
from $H$ (Eq.~(\ref{12p-rabimodel}))
as a function of the scaled coupling $g_1^{(\mathrm{eff})}/g_{1c}^{(\mathrm{eff})}$
for large values of $\Delta$ are shown in Fig.~\ref{Sz_Nph},
where $g_{1c}^{(\mathrm{eff})} = \sqrt{\Delta \omega^{(\mathrm{eff})}}/2$.
A negative $M$ emerges in the mixed QRM due to the positive effective bias.
Clearly, the photon number $N_{\mathrm{ph}}$ increases with the increase of $%
g_2$, which also indicates that the qubit-cavity interactions are enhanced
and more photons are excited. For different
two-photon coupling in the mixed QRM, the photons are considerably enhanced at
almost the same  scaled coupling around $1$.  Whether it is a signature of the
quantum phase transitions in the mixed QRM as observed in Ref.~\cite{peng2,ying} deserves further study.

\section{Summary}

In this paper, by using Bogoliubov operators, we exactly solve the mixed QRM
with both one- and two-photon terms analytically. The $G$-functions with the
pole structures are derived, which reproduce completely the regular spectra.
They can also be reduced to the unmixed ones. It is found that there are two
sets of poles associated with two Bogoliubov operators. Two types of
exceptional eigenvalues are then derived, which cannot be obtained solely by
requiring that the corresponding coefficients vanish like in the unmixed
models. When the two-photon coupling strength $g_2$ is close to $1/2$, two
collapse energies are derived. One is finite, while the other  diverges
negatively. All energy levels collapse to the lower one,
therefore diverge also, in sharp contrast to the unmixed two-photon model.
The level degeneracy in the unmixed model is relieved due to the absence of
parity symmetry. The avoided crossings are strictly discerned from the very
close levels in the mixed model by the absence of exceptional eigenvalues
around the "crossings".

We construct an effective one-photon Hamiltonian to describe the the mixed QRM, which is valid in weak two-photon coupling regime. The mixed QRM is equivalent to a single-photon
one with an effective positive bias, suppressed photon frequency and
enhanced one-photon coupling.   This feature in the mixed system is very
helpful to the recent circuit QED experiments where the intense competition
to increase one-photon coupling is performed in many groups~\cite%
{Niemczyk,exp,Yoshihara,Forn2}. We suggest that the simultaneous presence of
both one- and two-photon couplings would cooperate to provide richer physics.


\textbf{ACKNOWLEDGEMENTS}  This work is supported by the National Key Research and Development Program of China (No. 2017YFA0303002), the National Science Foundation of China (Grants No. 11674285 and No. 11834005).

$^{\ast }$ Corresponding author. Email: qhchen@zju.edu.cn

\end{document}